\documentclass{aastex631}
\usepackage{graphicx} 
\usepackage{verbatim}
\providecommand{\mum}{\mbox{$\mu$m}}
\providecommand{\thpp}{the previous processing}
\providecommand{\Thpp}{The previous processing}

\begin{document}

\author[0000-0003-0947-2824]{D.~R.\ Mizuno}
\affiliation{Institute for Scientific Research, Boston College, 140 Commonwealth Avenue, Chestnut Hill, MA 02467, USA}

\author[0000-0003-1955-8509]{T.~A.\ Kuchar}
\affiliation{Institute for Scientific Research, Boston College,
140 Commonwealth Avenue, Chestnut Hill, MA 02467, USA}

\author[0000-0002-2626-7155]{Kathleen E.\ Kraemer}
\affiliation{Institute for Scientific Research, Boston College, 140 Commonwealth Avenue, Chestnut Hill, MA 02467, USA}

\author[0000-0003-4520-1044]{G.~C.\ Sloan}
\affiliation{Space Telescope Science Institute, 3700 San Martin Drive, Baltimore, MD 21218, USA}
\affiliation{Department of Physics and Astronomy, University of North Carolina, Chapel Hill, NC 27599-3255, USA}

\author{Samantha Greene}
\affiliation{Institute for Scientific Research, Boston College,
140 Commonwealth Avenue, Chestnut Hill, MA 02467, USA}
\affiliation{Embry Riddle Aeronautical University, Prescott, AZ 86301, USA}

\author{Elianna\ Cohen}\affiliation{Institute for Scientific Research, Boston College,
140 Commonwealth Avenue, Chestnut Hill, MA 02467, USA}
\affiliation{Georgia Institute of Technology, Atlanta, GA 30332, USA}

\author{Holly Branco}\affiliation{Institute for Scientific Research, Boston College,
140 Commonwealth Avenue, Chestnut Hill, MA 02467, USA}

\received{2025 May 16}; \revised{2025 July 20} \accepted{ 2025 August 4}; 
\published{2025 September 23}

\title{An Improved Atlas of Full-Scan Spectra from ISO/SWS}

\begin{abstract}

We present an atlas of full-scan spectra from the Short-Wavelength Spectrometer (SWS) aboard the Infrared Space Observatory (ISO) after reprocessing and improving an earlier version published 22 years ago.  The SWS spectra cover the wavelength range from 2.35 to 45.3~\mum.  They include scans in 12 separate bands, and we have updated the methods used to combine those bands into a single continuous spectrum. The main improvement comes from applying multiple constraints, including new photometry and spectra from the Infrared Spectrograph (IRS) on the Spitzer Space Telescope that have become available since the release of the original products, and individualized attention to each spectrum, to renormalize the separate bands into a more consistent single spectrum. In particular this removed unphysical negative fluxes that were common in the original data products. The new database, with 1035 reprocessed spectra, will be available to the community at IRSA, which also hosts the original processing.   
\end{abstract}

\section{Introduction}\label{introduction} 

The Infrared Space Observatory \cite[ISO,][]{iso96} was a joint mission between the European Space Agency (ESA) and NASA that operated from 1995 to 1998.  Its four science instruments, a camera, a photo-polarimeter, and two spectrometers, obtained data from 2.4 to 240 \mum.  ISO was the first infrared space telescope focused on pointed observations instead of survey data.  It obtained over 26,000 science observations and over 4,000 calibration observations \citep{isohb03}.  Prominent discoveries include the 
first detections of interstellar HF, crystalline silicate features in asymptotic giant branch (AGB) stars, and water in the troposphere of Saturn, to name a few \cite[][respectively]{neufeldea97, watersea96, encrenazea99}.

The Short-Wavelength Spectrometer (SWS) on ISO was a scanning spectrometer with a wavelength range from 2.4 to 45 \mum.  For the full-scan mode (SWS01), spectra were obtained in all 12 bands, each with 12 detectors scanned up and down in wavelength \citep{sws96,swshb03}.  The result for a spectrum covering the full wavelength range was 288 spectral segments.  The average spectral resolving 
power $R$ was $\sim$600,\footnote{$R \equiv \lambda/\Delta\lambda$} with up to $\sim$2400 possible.  Roughly 1250 observations were taken in this mode, of which $\sim$1000 had a
detectable signal in the spectrum \citep{kspw02}.

 \cite{swsatlas,swsatlasdoi} developed an automated pipeline to reduce the SWS01 
 observations and generated an Atlas of uniformly reduced spectra for the community, available at NASA's Infrared Science Archive (IRSA).  They produced two versions of each spectrum. The first, designated {\em tdtnumber.pws}\footnote{TDT stood for ``target dedicated time'' and is a unique identifier number for each observation.} (hereafter referred to as ``pws'' data), provided the 12 segments (or bands) with overlapping data between adjacent segments intact and preserved any band-to-band discontinuities, and is the starting point for the new processing. The second, designated {\em tdtnumber.sws}, normalized the 12 segments to remove any discontinuities, and trimmed the overlapping data to produce a single, continuous spectrum. These results are referred to below as the ``previous'' processing, pipeline, and spectra.
 These data products have been widely used, with over 200 refereed citations and over 1000
 downloads\footnote{The Astrophysical Data System: ui.adsabs.harvard.edu}.

However, the previous pipeline made assumptions when adjusting spectral segments to remove discontinuities that could lead to flaws in the spectra and unphysical results, such as as negative fluxes.  We therefore undertook to revisit those steps. Section~\ref{swssample} briefly describes the SWS and the observed sample. Section~\ref{reprocessing} presents the revised processing method. Section~\ref{solution} discusses the additional constraints, solution, and quality assessment (QA) process, and Section~\ref{atlas} summarizes the results and describes the new Atlas.

\section{The SWS and the Observed Sample}\label{swssample} 

\subsection{The Sample}

The current sample is based on the SWS atlas \citep{swsatlas}, which included 1239 spectra of Galactic objects (many of the same source). Table~\ref{tab.sample} breaks this total down by the group numbers from the infrared spectral classifications by \cite{kspw02}.  Groups 1 to 5 include sources with increasingly red colors, with dust-free stars in Group 1, dusty stars in Group 2, spectra dominated by relatively warm dust (but little or no stellar continuum) in Group 3,
and red objects, primarily nebulae in star-forming regions or highly evolved objects like planetary nebulae, in Groups 4 and 5, with the reddest sources in Group 5.  Group 6 contains spectra with no significant continuum, and Group 7 consists of spectra with no detected flux or other flaws that prevented meaningful classification.

Table~\ref{tab.sample} gives the totals from each group in the original atlas and in this new spectral atlas.  All Group 7 spectra are excluded here. A number of spectra in Group 1 were intentionally mispointed for calibration purposes (14 of $\alpha$ Boo and 5 of $\gamma$ Dra) and are excluded because they cannot be corrected.  A handful of other spectra were excluded due to such problems as unintended mispointings (4 observations) or saturation (1 observation).\footnote{These numbers also differ slightly from those in Table 4 of \cite{kspw02} as the raw data for 9 observations were not available for the original processing to the pws level and are omitted from both the original atlas and this work.} Source variability is discussed in Section \ref{vars}.

\begin{deluxetable*}{rrrrc}
\tablecaption{SWS01 Sample\label{tab.sample}}
\tablewidth{0pt}
\tablehead{\colhead{Group} & \colhead{Original} & \colhead{This} & \multicolumn{2}{c}{Variable}\\
\colhead{Number} & \colhead{Atlas} & \colhead{Work} & \colhead{No.} &\colhead{\%}
}
\startdata
1 & 176 & 157 & 87 &  55\\
2 & 237 & 237 & 227 & 96\\
3 & 72 & 71 & 58 & 82 \\
4 & 219 & 217 & 109 & 50 \\
5 & 244 & 243 & 69 &28 \\
6 & 112 & 111 & 44 &40 \\
7 & 179 & 0 & \nodata& \nodata \\
Total & 1239 & 1035 & 594 & 57
\enddata
\end{deluxetable*}

\begin{figure} 
\centering
\includegraphics[scale=0.7]{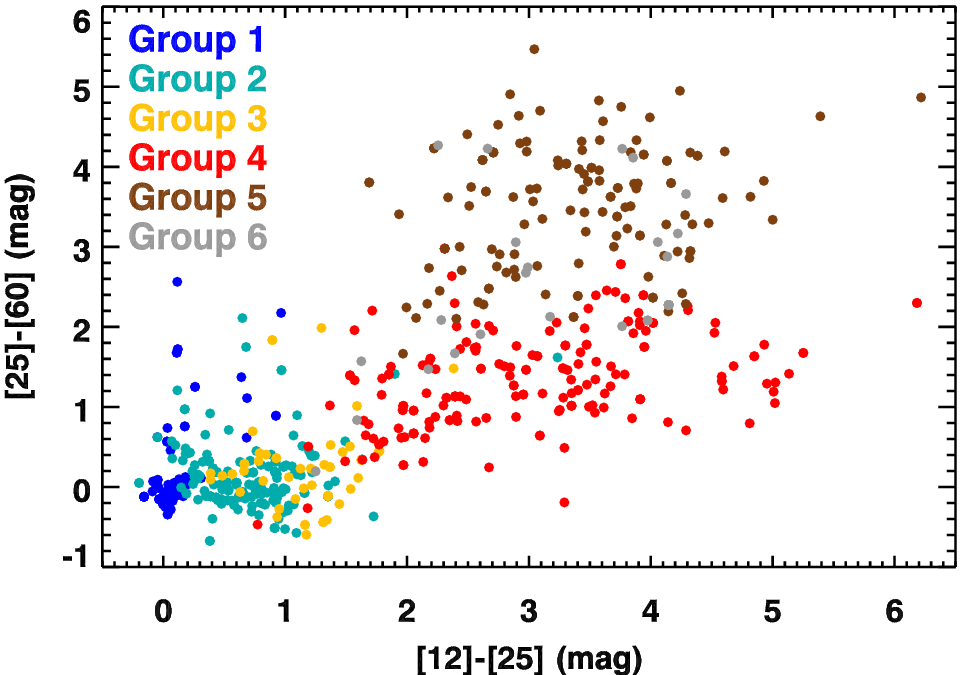}
\caption{Color-color diagram of the SWS sample using IRAS photometry. Symbol colors indicate the 6 groups.}
\label{fig.colcol}
\end{figure}

Figure~\ref{fig.colcol} plots the sample in color-color space using the flux densities from the Infrared Astronomical Satellite (IRAS) at 12, 25, and 60~\mum\ \citep{iras84, iraspscdoi,irascats88}. It shows how the
group classifications by \cite{kspw02} can be used to navigate the SWS
database.\footnote{This plot was introduced by \cite{vdv88} for evolved stars. Because
we have included the zero-magnitude flux densities in our color 
definitions, the [12]$-$[25] color is shifted by $+$1.56 and [25]$-$[60] 
by $+$1.88. Like them, we plot only the quality 3 data.}  The groups defined by \cite{kspw02} separate well, except 
that the bluer groups (1--3) tend to be more affected by stray light that reddens their [25]$-$[60] color.  The distinction in [25]$-$[60] is more physically 
meaningful for Groups 4 and 5.

\subsection{SWS01 Properties and the Original Processing}\label{swsproperties}

Table \ref{tab.swsbands} summarizes the characteristics of the 12 bands used in SWS01 observations. Scans were made at one of four speeds that determined the wavelength sampling and hence spectral resolving power ($R$), as well as the sensitivity of an observation. The spectral resolving power is for an extended source at Speed 1 and 2 near the band center \cite[their Tables 2.1 and 3.1]{swshb03}, as these were the most common scan speeds. Actual values of $R$ in the (raw) data also depended on the source type.  ``Full'' wavelength ranges include the overlaps between the segments. \citet[][their Table 2]{swsatlas} resampled the spectra onto a uniformly spaced wavelength grid. The same grid is used here, and the wavelength interval for each segment is given in the Wavelength Spacing column. The duration is the length of time for a Speed 1 scan in one direction \cite[see their Section 3]{swshb03}\footnote{Note there is no single integration time for an observation. The scan duration is governed by the scan speed and the wavelength range of a given band. This determines the spectral resolving power and the integration time for each band.}. We assign a segment number here to each band to facilitate use of the newly processed data files. Band 3E (segment 11) is omitted due to its high noise and generally unreliable behavior.

\begin{deluxetable*}{llccccc}
\tablecaption{SWS01 Characteristics\label{tab.swsbands}}
\tablewidth{0pt}
\tablehead{
\colhead{Band} & \colhead{Segment} & \colhead{Validated} & \colhead{Full} & 
\colhead{Nominal} & \colhead{Wavelength} & \colhead{Single Scan}\\
\colhead{Name} & \colhead{Number} & \colhead{Wavelengths} & \colhead{Coverage} & 
\colhead{Res. Power} & \colhead{Spacing} & \colhead{Duration}\\
\colhead{} & \colhead{} & \colhead{(\micron)} & \colhead{(\micron)} & 
\colhead{($R$=$\lambda/\Delta\lambda$)} & \colhead{(nm)} & \colhead{(sec)}
}
\startdata
1A & 1 & 2.38 -- 2.60 & 2.36 -- 2.64    & 500 & 0.50 & 70\\
1B & 2 & 2.60 -- 3.02 & 2.55 -- 3.05    & 400 & 0.67 & 70\\
1D & 3 & 3.02 -- 3.52 & 2.99 -- 3.78    & 490 & 0.67 & 170 \\
1E & 4 & 3.52 -- 4.08 & 2.99 -- 4.09    & 350 & 1.00 & 170\\
2A & 5 & 4.08 -- 5.30 & 4.01 -- 5.70    & 460 & 1.00 & 170\\
2B & 6 & 5.30 -- 7.00 & 5.22 -- 7.07    & 270 & 2.00 & 170\\
2C & 7 & 7.00 -- 12.0 & 6.93 -- 12.61   & 460 & 2.00 & 150\\
3A & 8 & 12.0 -- 16.5 & 11.89 -- 16.61  & 220 & 3.33 & 70\\
3C & 9 & 16.5 -- 19.5 & 15.97 -- 19.59  & 520 & 3.33 & 170\\
3D & 10 & 19.5 -- 27.5 & 19.28 -- 27.64 & 280 & 6.67 & 170\\
4 & 12 & 29.0 -- 45.2 & 27.50 -- 45.40  & 330 & 10.00 &150\\
\enddata
\tablecomments{Spectral resolving power ($R$) and  wavelength spacing are given for Speeds 1 and 2; scan duration is for Speed 1. See Sec. 2.2 for details.}
\end{deluxetable*}

As mentioned above, the mission-level pipeline provided data in 288 spectral segments, 24 for each of the 12 spectral bands. The original raw data were affected by a number of issues, such as errors in gains, biases, changes in dark levels during scans, and persistence. Band 2C was particularly susceptible to hysteresis, which could cause the scans up and down in wavelength to diverge significantly from each other.

\Thpp\ reduced and combined the 24 segments for each band, including outlier removal, median filtering, and resampling onto a uniform wavelength grid. Combining the up and down scans partially corrected for hysteresis effects in the detectors. The resulting pws files maintained the nonvalidated data in the overlap sections and any discontinuities between bands due to the instrument calibration.

\Thpp\ then aligned the 12 spectral bands by selecting one band as an anchor (either band 1E or 3D, depending on which had the higher mean flux density) and then adjusting each of the successive adjacent segments, with either an additive offset or a multiplicative scaling factor\footnote{Henceforth, scaling or scaling factor means a multiplicative adjustment and offset means an additive adjustment}, to match the mean flux value of the preceding segment in the mutual overlap region.  Scaling factors were the primary adjustment method. However, for bands with low fluxes, this commonly required large scaling factors to match the flux in the overlap region, thus greatly amplifying the noise.  Therefore, for bands with low fluxes, with medians below $\sim$20 Jy, the scaling was changed to an offset.

\begin{figure} 
\centering
\includegraphics[scale=0.7]{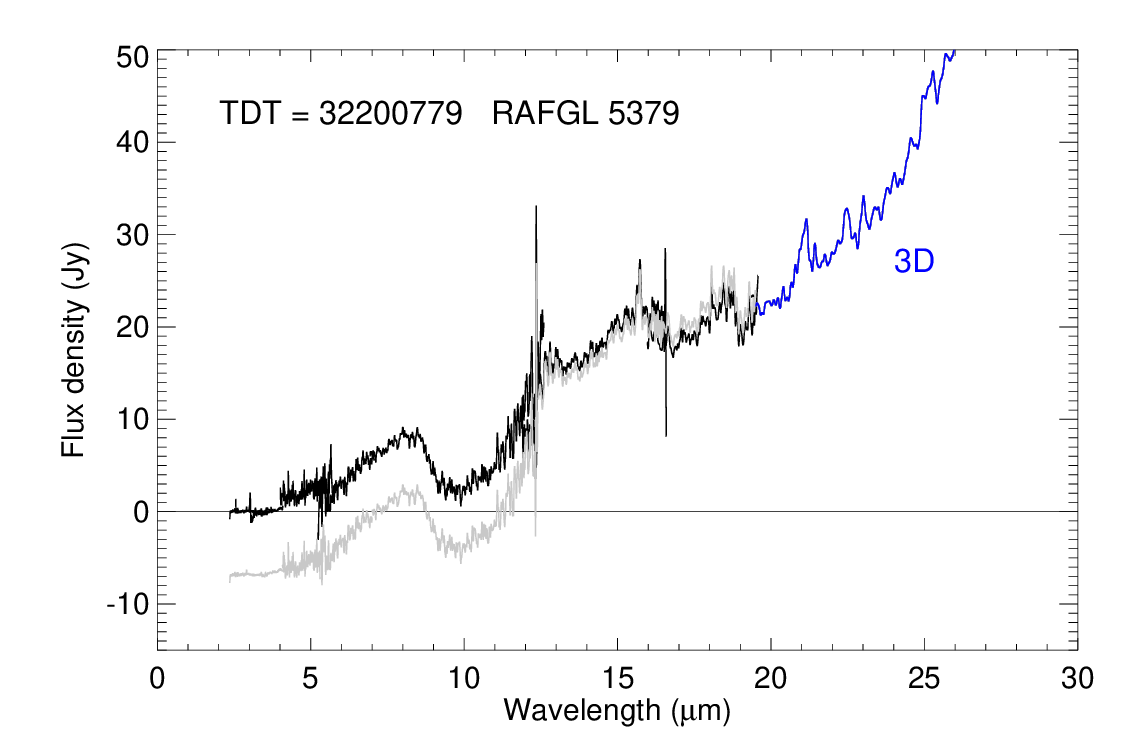}
\caption{Example of the normalization algorithm in \thpp\ driving the flux density negative. The original pws data are shown in black, and \thpp\ results are in gray. Band 3D was used as the anchor band, shown in blue. }
\label{fig.negflux}
\end{figure}

\begin{figure} 
\centering
\includegraphics[scale=0.7]{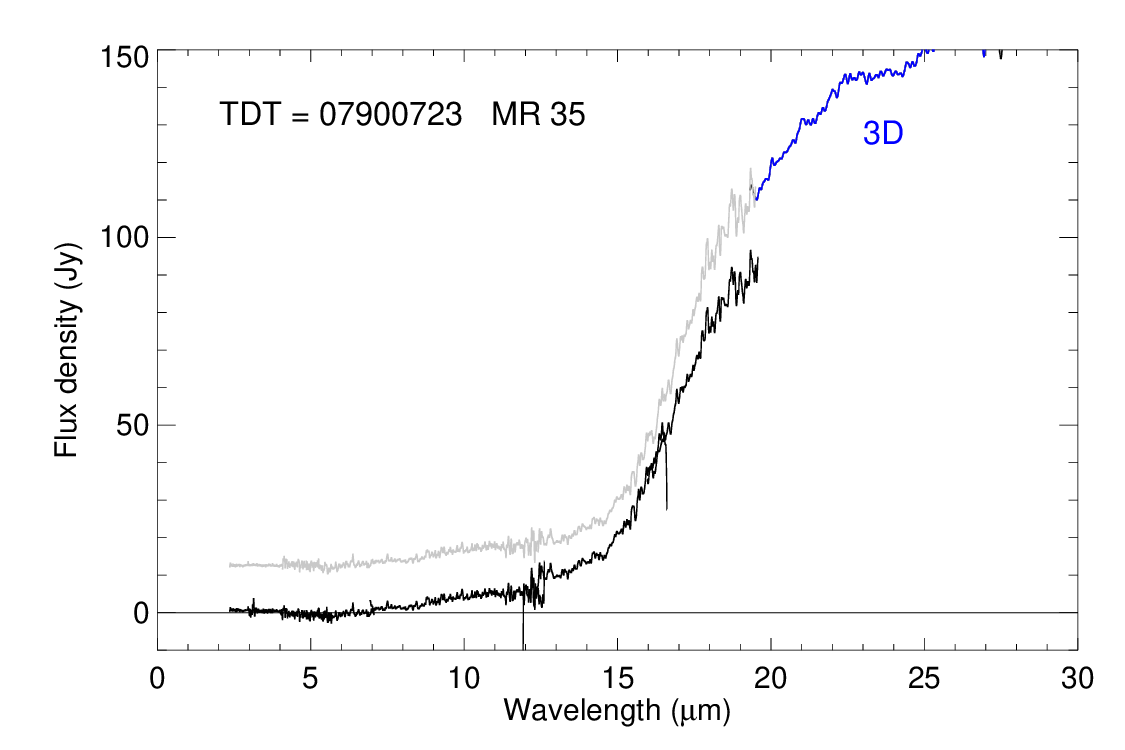}
\caption{Example of the normalization algorithm in \thpp\ apparently producing excess flux density at low flux levels. The color scheme is the same as for Figure \ref{fig.negflux}.}
\label{fig.excessflux}
\end{figure}

The assignment of a band with one of the higher flux densities as the sole anchor resulted in more distant bands in wavelength space being effectively untethered to any reference value.  In particular, the application of an offset for low-flux bands did not constrain the resulting adjusted fluxes of those bands.  In some cases the algorithm shifted low-flux bands to negative values.  Figure \ref{fig.negflux} shows an example.  The discontinuity at 12 \micron\ in the pws data, between bands 2C and 3A, and an application of an offset for the lower-wavelength bands resulted in negative values. In this case, applying a scaling factor for all bands would likely have given a reasonable result.  Even when the fluxes remain positive, the algorithm could introduce a considerable uncertainty to the low-level fluxes.  Figure \ref{fig.excessflux} shows an example of the algorithm apparently producing excessive flux at low levels. Approximately 150 (14\%) of the processed spectra had segments with negative resulting fluxes and another $\sim$240 (23\%) had markedly excess flux.

\section{Reprocessing method}\label{reprocessing} 

The algorithm in \thpp\ is ``exact'' in that the solution is fixed for a single selected anchor band, and so it does not allow for any additional constraints.  In the current work we allow multiple anchor bands, and also apply photometric values and spectra from Spitzer's Infrared Spectrograph \cite[IRS,][]{irs04} as additional constraints, as available.  These additional constraints require additional degrees of freedom to allow matching the endpoints of the segments while satisfying the added constraints.

Rather than attempting to add just enough degrees of freedom to allow a similar exact solution for any given set of constraints (i.e., by some scheme of adding additional scaling factors or offsets), a more general solution is to allow both an offset and a scaling factor for each spectral band and then optimize the parameters with a least-squares procedure.  The obvious issues with this approach are that the system will generally be under-determined, even with the additional constraints, and allowing both parameters for each segment permits implausible values to satisfy the constraints.  We address these issues by applying a further ``damped'' condition on the parameters such that the overall magnitude of the adjustments is also, in a sense, minimized in the least-squares solution.  This both permits a solution and generally restricts the parameter values to the minimum adjustment that will satisfy the constraints.  The second issue is also addressed by submitting the initial results to extensive QA (Sec. \ref{qasection}) in which the degrees of freedom may be restricted in a subsequent refit, or the constraints modified. 

We are not asserting that the offset and scaling factors resulting from the least-squares procedure are individually correct in any absolute sense.  The data do not provide enough information to determine such values.  The assignment of an offset and scaling factor for each segment is applied primarily as a method to increase the number of degrees of freedom to accommodate multiple additional constraints, with the assumption that these are each reasonable adjustments to make. 

\subsection{Least squares approach}

We allow each spectral segment $i$ to be adjusted with both an offset $\alpha$ and scaling factor (1 $-$ $\beta$), so that the adjusted segment flux density $F_{i}^{A}(\lambda)$ can be expressed

\begin{equation}
    F_{i}^{A}(\lambda) = \alpha_{i} + (1 - \beta_{i})F_{i}(\lambda),
\end{equation}

\noindent where $F_{i}(\lambda)$ is the intermediate pws stage of processing for band $i$.
The scaling factor is defined this way so that a zero scaling parameter $\beta$ corresponds to a null scaling adjustment.  We notate the complete set of parameters as

\begin{equation}
    \mathbf{x} = (\alpha_{1}, \beta_{1}, \alpha_{2}, \beta_{2}, . . . )
\end{equation}

\noindent for spectral bands labeled $1,2,....$  A standard least-squares approach is applied.  We define $\chi^{2}$ as a function of the parameters $\mathbf{x}$, and to be explicit and to show the notation used below, we have

\begin{equation}
    \frac{ \partial \chi^{2}} {\partial x_{n}} = 0
\end{equation}

\noindent for each parameter $x_{n}$, giving the system of equations (noting that the partial derivatives are linear in the coefficients $\mathbf{x}$, as shown in Appendix \ref{ref.chisq}),

\begin{equation}
    \mathbf{Ax} = \mathbf{B},
\end{equation}

\noindent where the elements of $\mathbf{A}$ and $\mathbf{B}$ are determined from Equation 3. Appendix \ref{ref.chisq} presents the contributions to $\chi^{2}$ from the band overlaps, damping factors, photometric values, and {\em Spitzer} IRS data.

\subsection{Uncertainties}

In \thpp, the reported uncertainties are the sample-to-sample scatter in the flux density, both for the original pws data and the normalized results. For the pws, that is the variance in the mean flux density at the re-gridded wavelength element from the 24 raw spectra. The uncertainties in the normalized spectral segments are the scatter in the original pws segment adjusted by the normalization scaling factor for that segment. Absolute photometric uncertainties are not provided, as the raw SWS data files did not include such information, and the normalization procedure did not provide any. The current processing generally has the same limitation. While the inclusion of photometric values and Spitzer IRS spectra in the least-squares normalization potentially allows a measure of the photometric accuracy for some of the spectra, for consistency over all the spectra we report uncertainties similarly to \thpp, with the reported values a measure of the scatter in the normalized spectral segments.

\section{Initial constraints, solution, and QA}\label{solution} 

The previous algorithm corrected for discontinuities between segments starting with a single anchor segment and propagating in both directions to the blue and red ends of the spectrum.  Any errors in the corrections would thus be compounded in both directions and in some cases could lead to unphysical results, such as negative fluxes.

We address this specific problem by assigning one or more additional anchors, specifically from among the band segments with the lowest positive median flux densities. Typically, the lowest positive segment is selected, although this may vary depending on the assessment of the data quality. In some cases, the lower-flux end of the spectrum is effectively anchored to either Spitzer IRS data or photometric values, without an explicit anchor segment selected.

\begin{figure} 
\centering
\includegraphics[scale=0.7]{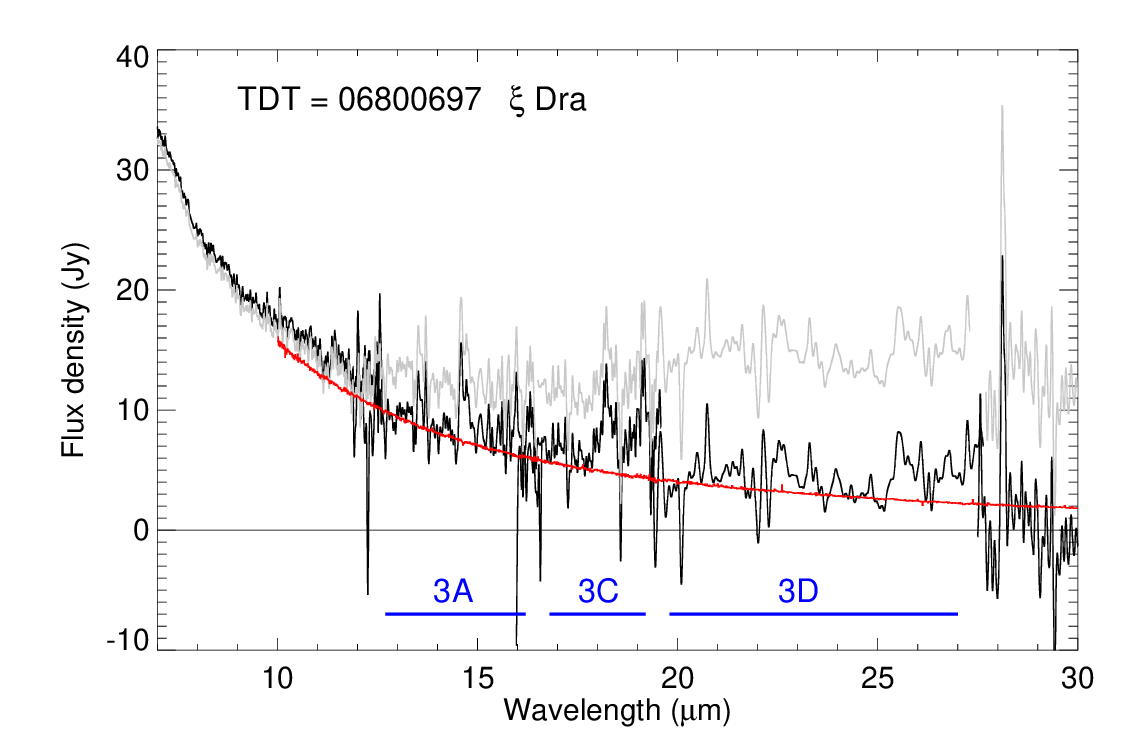}
\caption{Example of the low-flux portion of a stellar source. The pws data are shown in black, Spitzer IRS data in red, and the previous results in gray.  Wavelength ranges for selected bands are schematically marked with blue bars, with the bands labeled.}
\label{fig.lowflux}
\end{figure}

One issue is the expected reliability of low-flux segments in the pws data. 
Figure \ref{fig.lowflux} shows a stellar spectrum for which we have Spitzer IRS data.  The plot includes the pws data, the IRS spectrum, and the previous results for $\xi$ Dra as an example. In this case the unadjusted pws data agree reasonably well with the IRS data, notably for bands 3A, 3C, and 3D. These results are typical for stellar sources for which IRS data are available, particularly in cases such as this where several adjacent band segments have mutually consistent flux levels in the pws data.

The initial corrections are made with the selected anchors and inclusion of any available photometric values and IRS data.  An additional initial condition is that band segments that have a negative median in the pws data are given an offset to achieve a zero median, and then participate normally in the fitting.  For the solution, the matrix ${\mathbf A}$ and vector ${\mathbf B}$ are determined from $\chi^{2}$ (Appendix \ref{ref.chisq}). Anchor segments are implemented by including them in the $\chi^{2}$ calculation,
then deleting that segment's corresponding rows and columns from ${\mathbf A}$ and ${\mathbf B}$.  The solution parameters ${\mathbf x}$ are determined from

\begin{equation}
    {\mathbf x} = {\mathbf A}^{-1}{\mathbf B}
\end{equation}

\noindent using the IDL\textsuperscript{\textregistered} function {\bf invert}.

\subsection{Quality assessment (QA)}\label{qasection} 

The QA for the resulting spectrum for each source generally involves two steps.  First, at least two authors visually searched for improper evaluations of the fluxes in the band overlap regions, due to poor-quality data.  Section \ref{overlaps} discusses the available remedies.  Second, the fitting parameters are examined for general segment-to-segment consistency, in particular the continuity of the slope across band boundaries.

\begin{figure} 
\centering
\includegraphics[scale=0.7]{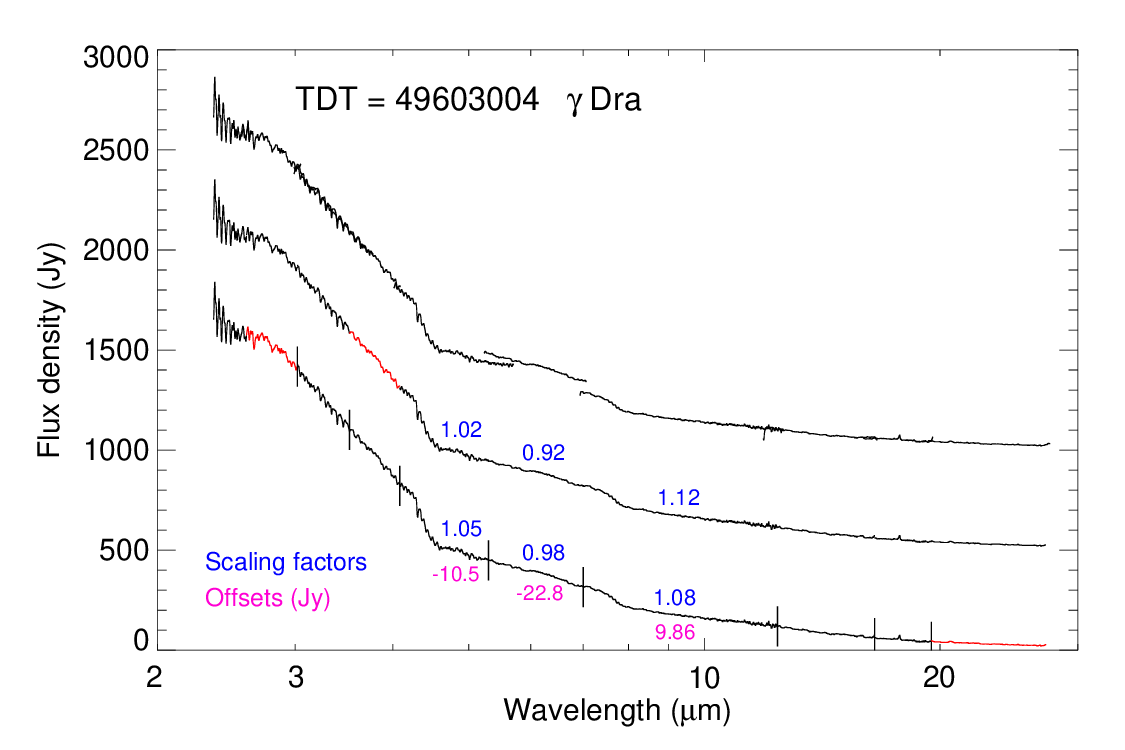}
\caption{Example of a common case in which the previous algorithm yielded good results, and the least-squares method produces essentially equivalent results.  Top: the original pws data; middle:  the previous results; bottom:  the result of the least-squares fitting.  The upper two plots have been vertically offset for clarity.  The anchors applied are shown in red.  The vertical bars show the band segment boundaries, and the determined corrections are shown for selected bands, scaling factors above the plot (in blue), and offsets for the least-squares fit below the plot (in magenta).}
\label{fig.basicfit}
\end{figure}

Figure \ref{fig.basicfit} shows a common example for which the previous algorithm yielded a good result, and the least-squares algorithm produced an essentially equivalent result.  In this case, a single additional spectral segment is used a second anchor, with no other constraints applied.  The top plot shows the original pws data, the middle plot shows the previous results, and the bottom plot shows the results of the least-squares fitting. In the latter two cases, the anchor segments used are shown in red. The determined corrections are shown on the plots for selected band segments.  This result demonstrates the utility of the new method.  It produces a result similar to the previous result when the data are more reliable.

\begin{figure} 
\centering
\includegraphics[scale=0.7]{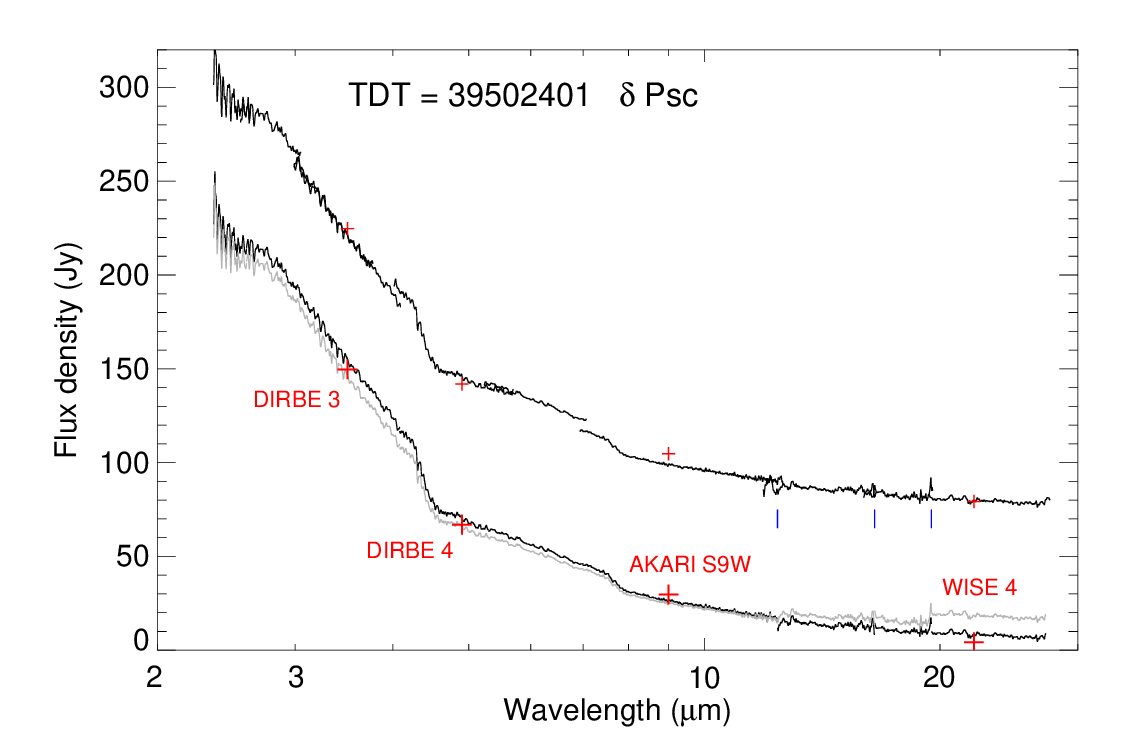}
\caption{Example of including photometric values in the least-squares fit.  The upper plot is the original pws data (offset for clarity), the lower plot is the least-squares result (in black) and the previous result (in gray).  The photometric values are show as red crosses.  The blue hash marks show where extrapolation was used to determine flux levels in the overlaps to compensate for noisy data.}
\label{fig.photofit}
\end{figure}

Figure \ref{fig.photofit} shows an example of including photometry values in the least-squares fit.  In this case the photometry is the only additional constraint; no anchor segments are applied. The upper plot is the original pws data, and the lower plot is the least-squares result, with the previous processing in gray. This case also shows an example where linear extrapolation was used to determine flux levels in selected overlaps (Appendix \ref{overlaps}); these overlaps are indicated with the blue hash marks. Noisy data in these overlaps caused discontinuities in the flux density levels in the previous results. These problems are identified, and the extrapolation remedy selected, as part of the QA process. 

In \thpp, the scaling factors, being the sole adjustment (above a certain flux level), are constrained by the separation in flux density between the spectral segments in the overlap regions and generally fall in the range 0.8 to 1.2.  Allowing both a multiplicative and additive correction applies no such effective constraints and could lead to unrealistic solutions. Problematic scaling factors are generally those below $\sim$0.75 or above $\sim$1.25, or when values between adjacent band segments are too noisy, in particular when the noise causes discontinuities in the slope of the resulting spectrum at the boundaries between bands. Remedies for problematic scaling factors include limiting those band segments to only offsets or only scaling factors, or for cases of excessive band-to-band scatter, applying a process we refer to as ``ganging.''  Ganging has two options: (1) a set of adjacent segments is given the same scaling factor while the offsets vary; or (2) the same offset is applied while the scaling factors vary (as restricting the offsets also effectively restricts the scaling factors).

\begin{figure} 
\centering
\includegraphics[scale=0.7]{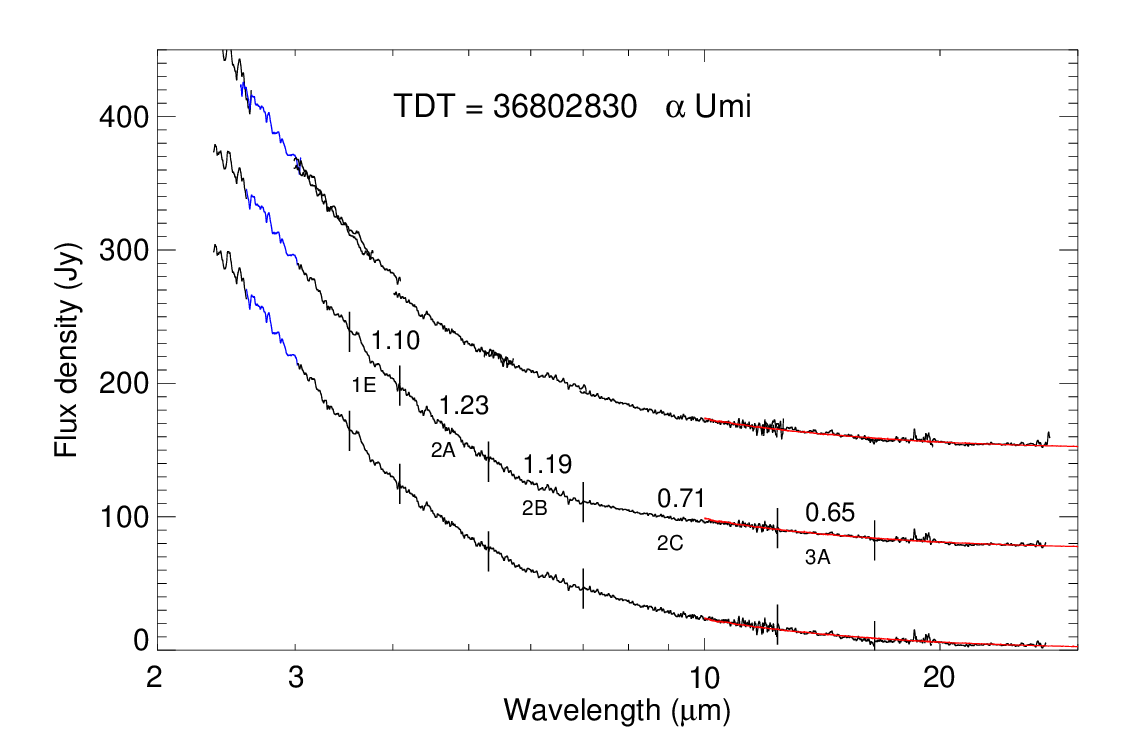}
\caption{Example of ganging scaling factors to correct a discontinuity in the slope in the initial fit (see text). The top plot is the original pws data, and the middle plot is the initial least-squares fit with the resulting scaling factors shown for several contiguous segments. The anchor segment (band 1B) is shown in blue, and the IRS data, included in the fit, are shown in red. The bottom plot shows the results of ganging the scaling factor for these five segments, with a value of 1.09. Spectra are offset for clarity.}
\label{fig.gang}
\end{figure}

Figure \ref{fig.gang} shows an example of this kind of correction. The scatter in the scaling factors causes a modest but clear discontinuity in the slope between segments 2B and 2C.  The bottom plot shows the refit with the scaling factors for segments 1E to 3A ganged together.  Slope discontinuities of this nature also occur occasionally in the previous results, so this is not a problem specific to the current processing.

A third subject of the QA is the effectiveness of including available photometric values in the fitting.  First, photometry values, and IRS spectra, are generally excluded from variable sources because the epochs of the ISO data and the various photometric values can differ significantly. Second, including photometry with low signal-to-noise ratios in the fit can create distortions in the resulting spectrum, evidenced by excessive scaling factors, slope discontinuities, or occasional residual gaps in the band overlaps.  These problems are remedied by removing one or more photometric values from the fitting.

An additional option for cases where the photometry leads to unacceptable distortions in the resulting spectrum is to apply the photometry after the initial fit.  In this procedure, the fitting is first performed with the photometry omitted.  Then, the resulting spectrum is given a global scaling factor to match the photometry in a least-squares sense, with a $\chi^{2}$ analysis similar to Section \ref{ref.photom} but with a single scaling parameter applied.  

\subsection{Repeatability issues}

As noted in Section~\ref{swsproperties}, the raw SWS data contained a number of issues that prevented effective, uniform calibration, which could not be solved satisfactorily so long after the end of the ISO mission. We focused on improving the post-detector calibration, which involved aligning the band segments to a continuous spectrum and applying additional constraints as appropriate both to match existing photometric values and avoid unrealistic flux levels. This effort required a degree of subjectivity, to evaluate and compensate for the generally poorly calibrated data in the overlap regions of adjacent segments, and to select additional constraints. We attempted to minimize the subjectivity by having multiple co-authors independently perform the same steps with the same criteria. While this process is not fully objective, we are confident that tying the calibration of the SWS data to photometry where possible, and mitigating for unphysical behavior, greatly improves on the previous calibration.

We note that the reduction approach adopted for the new atlas requires personal attention to many individual spectra, which is manageable for a database of $\sim$1,000 spectra.  For significantly larger databases, much of the decision-making would have to be automated.

\section{Results}\label{atlas} 

The reprocessing has resulted in an improved database with 1,036 SWS spectra for the community to use. Below, we discuss the contents of the new atlas along with some of the trends in the data.

\subsection{Trends} 

We assessed the results for possible influence of observational parameters or source characteristics on the derived corrections. The orbit number can be used as a proxy for the mission length and could potentially reveal sensitivity changes that were uncorrected by the instrument pipeline. The average offsets and scaling factors, though, do not measurably change from the beginning of the mission to its end. Differences among the 6 spectral groups could indicate uncorrected dependencies on the source color (i.e., spectral energy distribution). Again, the corrections do not differ for the blue sources and the red sources. We do find that the corrections between the four segments of Band 1 were consistently smaller (averaging $\sim$1--3\%) than those between other segments (averaging $\sim$5--10\%). This can be attributed to the generally better behavior of InSb detectors (Band 1) compared to Si:Ga (Band 2) or Si:As (Bands 3 and 4) detectors that were available to the mission.
  
\subsection{Variability}\label{vars} 

Table~\ref{tab.sample} includes the number and percentage of variables 
observed in each group. Known variables were identified using the General Catalogue 
of Variable Stars \citep{gcvs17}, the IRAS Point Source Catalog \citep[PSC;][]{irascats88}, and the DIRBE post-cryogenic catalog \citep{priceea10}.
For the IRAS PSC, stars with a variability index $\ge$ 35  were identified
as variable (i.e., $\ge$35\%\ likelihood).  The vast majority of Groups 2 (96\%) and 3 (82\%) are variables,
because most of the stars in these groups are supergiants or on the 
asymptotic giant branch, undergoing pulsations as long-period 
variables, and forming dust in their outflows \citep[e.g.,][]{hab96,
kspw02}.  More than half of the stars in Group 1 are also variable,
indicating that while this group does not include dust-producing stars,
it is dominated by stars not on the main sequence.

Overall, more than 50\% of the SWS01 observations were of variable sources. 
None of the available photometry was taken contemporaneously with the spectra. This lack of simultaneity limited our use of the photometry to sources not known to be variables. Thus, in addition to the formal instrumental uncertainty in the absolute flux densities \cite[1$\sigma$$\sim$4--22\%;][]{swshb03}, users should bear in mind the intrinsic variability of most SWS01 sources.

\subsection{The New Atlas} 

The reprocessed spectra will be in Infrared Processing and Analysis Center (IPAC) table format. For each spectrum, there are two files.  The first contains the adjusted band flux densities, with the overlaps trimmed to produce a single continuous spectrum, using the band boundaries from \cite{swsatlas}.  The table contains five columns, with one entry for each wavelength sample.  The columns are: the wavelength of the sample in microns, the adjusted flux density in Jy, an error estimate in Jy, the segment number, and a flag for whether the sample is in an overlap region with a neighboring band. The IPAC table file header contains information about the observation, the band-by-band adjustment parameters, and other processing information.  These files are designated \textit{tdtnumber\_newsws.tbl}. Appendix~\ref{header} describes the contents of the header.

The second table file contains both the original pws band data and the adjusted data, without the overlap trimming. The header information is identical. This file contains seven columns: the wavelength in microns, the adjusted flux density in Jy, an error estimate of the flux density in Jy, the segment number, the original pws flux density in Jy, an error estimate of the pws flux density in Jy, the band index, and the overlap flag. These files are designated \textit{tdtnumber\_newpws.tbl}.

The database will be hosted at IRSA\footnote{irsa.ipac.caltech.edu} and will also be available on the Boston College node of the Dataverse\footnote{dataverse.harvard.edu/dataverse/bc}.

\begin{acknowledgements} This work is based on observations made with the Infrared Space Observatory (ISO), an ESA project with instruments funded by ESA Member States (especially the PI countries: France, Germany, the Netherlands and the United Kingdom) and with the participation of ISAS and NASA.  Financial support was provided by the NASA ADAP through grant 80NSSC21K0985. We thank the anonymous referee whose comments helped clarify the paper.
This research used the NASA/IPAC Infrared Science Archive, which is funded by the National Aeronautics and Space Administration (NASA) and operated by the California Institute of Technology, NASA’s Astrophysics Data System, and CDS’s Vizier and Simbad services.  
\end{acknowledgements}

\bibliographystyle{aasjournal}
\bibliography{refs.bib}

\appendix

\section{The definition of $\chi^{2}$}\label{ref.chisq}

The total $\chi^{2}$ is the sum of the contributions from the band overlap analysis, the photometric values, the Spitzer IRS spectra, and a damping procedure:

\begin{equation}
    \chi^{2} = \chi^{2}_{OV} + \chi^{2}_{PHOT} + \chi^{2}_{IRS} + \chi^{2}_{DAMP}
\end{equation}

\subsection{Band overlaps}\label{ref.overlaps}

With equation (1), the contribution to $\chi^{2}$ of the overlap of spectral segments $i$ and $j$ is defined as

\begin{equation}
\chi^{2}_{[ij]}  =  \sum_{\lambda \in [ij]}[F_{i}^{A}(\lambda) - F_{j}^{A}(\lambda)]^{2}
\end{equation}
\begin{equation}
  =  \sum_{\lambda \in [ij]}[\alpha_{i} + (1 - \beta_{i})F_{i}(\lambda) - \alpha_{j} - (1 - \beta_{j})F_{j}(\lambda)]^{2},
\end{equation}
\noindent where the summation depends on how the differences in the overlap regions are calculated. As a default, the flux values of one of the bands are interpolated to the wavelength values of the other band, and the summation is over the resulting wavelength samples.  Here, we are using the notation $[ij]$ to represent the overlap region between bands $i$ and $j$.
The total contribution of the band overlaps to $\chi^{2}$ is then

\begin{equation}
\chi^{2}_{OV}  =  \sum_{[ij]}\chi^{2}_{[ij]}.
\end{equation}

\subsubsection{Evaluation of the overlaps}\label{overlaps}

\begin{figure} 
\centering
\includegraphics[scale=0.7]{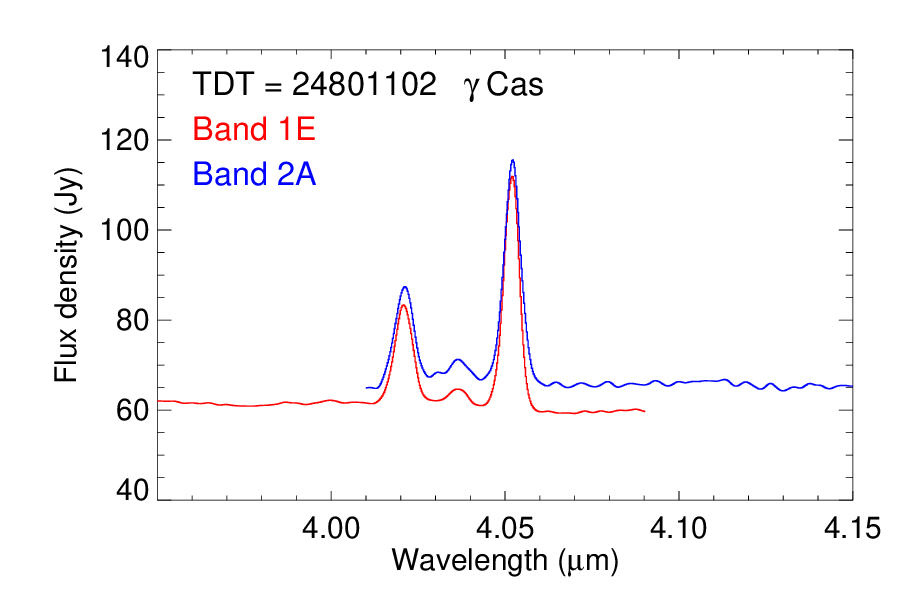}
\caption{Spectral features in a band overlap region that could in principle influence the relative scaling factor between the two bands.  We delete the 4.05~\mum\ features from the overlap analysis as \cite{swsatlas} did.}
\label{fig.overlap}
\end{figure}

In \thpp, the overlap differences were evaluated using the average flux density of each of the bands over the specified wavelength range of the overlaps.  In this work, as stated above, the default method is to preserve the sample-by-sample differences after applying the interpolation to equalize the wavelength sampling.  In principle, this allows the resulting relative scaling factors for the two bands to be correlated in the least-squares solution if the overlap region includes spectral features (Figure \ref{fig.overlap}).  However, this situation is fairly rare and the data quality in the overlap regions is likely not sufficient to provide any actual benefit.  For the general case of uncorrelated noise in the overlaps, this procedure should not differ materially from using the overlap averages as in \thpp, except for the effect on the relative weighting of the overlaps in $\chi^{2}$, discussed below.  Also, it should be noted that including the sample-by-sample differences in $\chi^{2}$ could potentially favor scaling factors less than 1 in the solution, because scaling downward does reduce the RMS of the sample differences.  The application of additional constraints on the solution beyond the overlap-matching itself mitigates this potential issue.

Several alternatives to this default procedure were used to address problems apparent upon inspection of the initial results in the QA process, typically due to poor-quality data in the overlaps.  The first is to replace the sample values in the overlap regions with either the average or the median of each band in the overlap (the sampling is preserved to allow the relative weighting of the overlaps to be assigned separately).  Applying the median, in particular, mitigates the effect of noise spikes and other isolated bad data samples in the overlap regions.  However, the selection of the average or median is applied over all the overlaps for a given spectrum and not band-by-band.

Second, for band segments with extremely poor quality, particularly when nearly all the data in the overlaps have anomalous values, the option of using the overall median of the band segment as the overlap value is allowed.  In effect, this removes the segment from participation in the fit except to provide a scalar level for the flux density to which the adjacent segments are normalized. In some cases, where even this procedure does not produce reasonable results, {\em ad hoc} offsets are applied as part of the QA process to align the segments.

Third, for less extreme cases of band segments with poor quality in the overlap regions, the option of applying a polynomial extrapolation from the ``good'' portion of the data into the overlap region is allowed.  The wavelength region over which to determine the polynomial, and the degree (usually linear) are selected (for a given band segment and overlap region), and the resulting polynomial function is evaluated at the samples in the overlap region for that band and overlap, which are then used for the $\chi^{2}$ contribution. 

Lastly, \thpp\ omitted band 3E from the end-to-end matching, due to poor quality data, and the neighboring bands 3D and 4 were matched to each other using an extrapolation method that depended on the source type.  We have also omitted band 3E from the fitting procedure, and use the same extrapolation methods to evaluate the ``overlap'' between bands 3D and 4.

\subsection{Damping factors}

With both an offset and a scaling factor allowed for all band segments, even with additional constraints beyond a single anchor segment, such as multiple anchors,  matching to photometric values or Spitzer IRS data (described below), the parameters will still in general be under-determined.  To remedy this problem, we apply an additional constraint that includes the overall magnitude of the adjustment parameters in the minimization.  To $\chi^{2}$ we add the contribution:

\begin{equation}
    \chi^{2}_{DAMP} = \sum_{n} C_{n}x_{n}^{2},
\end{equation}

\noindent where the sum is over the parameters $n$, and the $C_{n}$ values are arbitrary coefficients. The coefficients are set small enough to allow the minimization conditions of the other $\chi^{2}$ contributions, particularly the band overlaps, to be nearly unchanged; but large enough to avoid the singularity condition of the resulting least-squares solution matrix $\mathbf{A}$, to within the precision of the computation.

These terms will only contribute to the diagonals of $\mathbf{A}$, e. g., for $x_{n}$ we have
\begin{equation}
\frac{\partial \chi^{2}_{DAMP}}{\partial x_{n}} = 2C_{n}x_{n},
\end{equation}

\noindent and so $2C_{n}$ will be added to the diagonal element $A_{nn}$.  We can pick the $C_{n}$ to scale the diagonal element by some small quantity $\epsilon$:
\begin{equation}
    2C_{n} = \epsilon A_{nn},
\end{equation}
\begin{equation}
    A_{nn}  \Rightarrow  (1 + \epsilon)A_{nn} 
\end{equation}

This step allows us to select a suitably small $\epsilon$ that will provide a reasonably uniform damping effect for all the parameters regardless of the scale of the matrix elements.  Empirically, we find that a value $\epsilon$ = 1$\times$10$^{-3}$ gives stable solutions and has a negligible effect on the flux-matching conditions for the band overlaps.  In principle, we could use differing values of $\epsilon$ to favor, for example, scaling factors over offsets, either globally or for individual band segments, but we have applied a single $\epsilon$ for all parameters.

\begin{deluxetable*}{lcccccc}
\tablecolumns{7}
\tablewidth{0pt}
\tablecaption{Photometric Filters and SWS Coverage}
\tablehead{
    \multicolumn{2}{c}{} & \multicolumn{3}{c}{Wavelength ($\mu$m)} & \colhead{Saturation} & \colhead{SWS} \\
    \colhead{Mission}&\colhead{Band} & \colhead{Effective} &
    \colhead{Min} & \colhead{Max} & \colhead{limit (Jy)} &
    \colhead{coverage\tablenotemark{a}}
}
\startdata
AKARI & S9W & 9.0 & 5.7 & 12.5 & $\gtrsim$500 & 63\%\\
      & L18W & 18.0 & 13.5 & 28.7 & $\gtrsim$300 \\
\cline{1-7}
DIRBE & Band 3 & 3.5 & 2.95 & 4.23 & - & 42\% \\
      & Band 4 & 4.9 & 4.32 & 5.35 & - \\
      & Band 5 & 12.0 & 7.7 & 17.5 & - \\
      & Band 6 & 25.0 & 15.5 & 29.3 & - \\
\cline{1-7}
WISE  & Band 4 & 22.2 & 19.4 & 28.0 & 12.0  & 25\%\\
\cline{1-7}
IRAS  & 12 & 12.0 & 7.0 & 15.5 & - & 57\% \\
      & 25 & 25.0 & 16.0 & 31.5 & - \\ 
\enddata
\tablecomments{See Section \ref{ref.photom} for details.}
\end{deluxetable*}\label{phototable}

\subsection{Photometric values}\label{ref.photom}

Where available, we have used photometric values from selected filters of the AKARI \citep{akari07,akariirc10, akariircdoi}, DIRBE \citep{dirbe98,dirbepsc04, dirbepscdoi}, IRAS \citep{iras84, iraspscdoi, irascats88} and WISE \citep{wise10, allwisedoi} missions to aid in the adjustment of the ISO band segments. Table \ref{phototable} shows the filters used in the processing. The last column gives the fraction of SWS observations which had photometry within 10\arcsec\ and S/N$>$2.5 (in any band), although the amount of usable photometry was smaller. As noted above, we did not use the photometry of known variable stars since the SWS and the photometric measurements were not simultaneous (Table \ref{tab.sample} and Sec. \ref{vars}). Similarly, while $\sim$17\% of the SWS observations had an IRS spectrum, a smaller fraction were usable due to variability.

In general, a signal $dn$ measured from a source with spectral flux density $F_{\lambda}(\lambda)$ can be expressed
\begin{equation}
    dn \propto \int F_{\lambda}(\lambda)\frac{R(\lambda)}{h\nu}d\lambda,
\end{equation}

\noindent where $R(\lambda)$ is the responsivity of the filter in units of electrons/photon, and the integration is over all positive values of $R$.

For AKARI, DIRBE, and IRAS, the reported flux density $F_{\lambda}^{*}$ of a given source is defined as the flux density at a specified reference wavelength $\lambda_{0}$ of a $1/\lambda$ flux density distribution (i.e., $F_{\lambda}^{*}\lambda_{0}/\lambda$) that produces the same signal, so we can equate

\begin{equation}
\int F_{\lambda}(\lambda)\frac{R(\lambda)}{h\nu}d\lambda = \int \frac{F_{\lambda}^{*}\lambda_{0}}{\lambda} \frac{R(\lambda)}{h\nu} d\lambda.
\end{equation}

Solving for $F_{\lambda}^{*}$ gives

\begin{equation}
F_{\lambda}^{*} = \frac{\int F_{\lambda}(\lambda)R(\lambda)\lambda d\lambda}{\lambda_{0}\int R(\lambda)d\lambda},
\end{equation}

\noindent and converting to units of $F_{\nu}$:

\begin{equation}
F_{\nu}^{*} =  \frac {\lambda_{0}\int \frac{F_{\nu}(\lambda)}{\lambda}R(\lambda)d\lambda}{\int R(\lambda)d\lambda}.
\end{equation}

For WISE, the reported flux densities are defined such that a {\em constant} $F_{\lambda}(\lambda)$ gives the same signal as the source, so a similar derivation gives

\begin{equation}
   F_{\nu}^{*} =  \frac {\lambda_{0}^{2}\int \frac{F_{\nu}(\lambda)}{\lambda}R(\lambda)d\lambda}{\int R(\lambda)\lambda d\lambda} 
\end{equation}

\noindent and noting also that the WISE fluxes are reported as magnitudes, and we are using the indicated zero-magnitude fluxes to convert to $F_{\nu}$.

In our case, for the synthetic photometry, $F_{\nu}(\lambda)$ is the set of adjusted band segments $F_{i}^{A}$, and a given filter $R(\lambda)$ may subtend multiple segments, so the $\chi^{2}$ contribution for a single photometric value $F_{\nu}^{K}$ (for the AKARI, DIRBE, and IRAS filters) is defined as

\begin{equation}
    \chi^{2}_{K} =
        \frac{1}
        {\sigma_{K}^{2}}
        \left(
            R_{K}^{*}
            \sum_{i} 
            \left(
                \int \left[ \alpha_{i} + (1 - \beta_{i})F_{i}(\lambda)\right] \frac{R_{K}(\lambda)}
                          {\lambda}d\lambda
            \right) - F_{\nu}^{K}
       \right)^{2},
\end{equation}

\noindent with

\begin{equation}
R_{K}^{*} \equiv \frac{\lambda_{0}}{\int R_{K}(\lambda) d\lambda}, \label{eqn.filter-SWS}
\end{equation}

\noindent where the summation is over all SWS bands covered by the filter $R_{K}$, the integration is over the portion of each SWS band $i$ subtended by the filter (the boundaries between bands are from \cite{swsatlas}, their Table 1), and $\sigma_{K}$ is the reported fractional uncertainty in $F_{\nu}^{K}$ applied as a relative weighting factor for each photometric value (the overall weighting for the photometric sources is described below). An identical expression holds for the WISE filters, with

\begin{equation}
    R_{K}^{*} \equiv \frac{\lambda_{0}^{2}}{\int R_{K}(\lambda)\lambda d\lambda}. \label{eqn.filter-WISE}
\end{equation}

For the integrations, the $R_{K}(\lambda)$ sampling is interpolated to the ISO wavelength samples for the given band, and a simple trapezoid rule is used. 

The $\chi^{2}$ contribution of the photometric sources is then

\begin{equation}
\chi^{2}_{PHOT} = \sum_{K} \chi^2_{K}.
\end{equation}

\subsubsection{Inclusion of source photometry}

For equations \ref{eqn.filter-SWS} or \ref{eqn.filter-WISE} to hold, the responsivity function $R(\lambda)$ must be completely covered by SWS data, so as a first criterion only those filters are included.  Second, an upper limit to the fractional error $\sigma_{K}$ is applied; for most cases a limit of 0.3 is set.  Third, some filters can saturate at the flux densities typical in the SWS data (in particular WISE W4); sources near or above the saturation limit are omitted.  Fourth, for individual cases, sources may be omitted in the QA process after the initial fitting if the photometric sources distort the shape of the spectrum (e.g., producing discontinuities in the slope at the band boundaries or residual gaps in the band overlaps) likely due to errors in the reported values (and noting that we are including sources with significant fractional uncertainties, albeit with relatively low weighting).  Also, for sources known to be variable, the photometry is usually omitted entirely to avoid including data from varying epochs.  Likewise, for extended sources the photometry is also usually omitted due to the effects of different aperture sizes. 

\subsection{Spitzer IRS spectra}

The Spitzer IRS spectra \citep{cassis15} are supplied as $F_{\nu}^{IRS}(\lambda)$, so we define the $\chi^{2}$ contribution for a single spectrum:

\begin{equation}
\chi_{IRS}^{2} = \sum_{i} \sum_{\lambda \in i} [F_{i}^{A}(\lambda) - F_{\nu}^{IRS}(\lambda)]^{2}
\end{equation}

\begin{equation}
= \sum_{i} \sum_{\lambda \in i} \left\{[\alpha_{i} + (1 - \beta_{i})F_{i}(\lambda)] - F_{\nu}^{IRS}(\lambda)\right\}^{2}.
\end{equation}

\noindent The SWS wavelengths are interpolated to the IRS sampling, and the inner summation is over the resulting wavelength values for a given SWS band. The outer summation is over the SWS bands $i$ subtended by the IRS spectrum.  As before, the boundaries between the bands are from \cite{swsatlas}.  The contributions to $\chi^{2}$ are sample-by-sample so it is not necessary for the IRS spectrum to be completely covered by the SWS bands. 

\subsection{Weighting}

The relative weighting of the band overlap contributions $\chi^{2}_{[ij]}$ depends on the objectives of the segment-matching effort. The matching algorithm used in \thpp\ did not allow the uncertainty of the flux-density averages determined in the overlap regions to influence the results.
In the current work, however, minimizing the level differences in the band overlaps is in principle competing with other constraints, so the choice of weighting can affect the desired results.  Noting that the quality of the results is evaluated by the global characteristics of the resulting spectrum and not whether the offsets and scaling factors are individually correct in a maximum-likelihood sense, the weighting assigned is largely an empirical choice.      

The default weighting of the band overlaps is proportional to the number of data samples in the overlaps.  Formally the weighting should also be scaled by $1/(\sigma_{i}^{2} + \sigma_{j}^{2})$, where $\sigma_{i}$ and $\sigma_{j}$ are the per-sample uncertainties in bands $i$ and $j$ in the mutual overlap region.  The potential issue is that with weighting by sample number and noise, overlaps with comparatively few samples or very noisy data could be sacrificed in the least-squares solution if other local constraints with higher weighting can be met, possibly leaving mismatches in the levels.  Alternatively, the overlaps can be given equal weighting to better ensure level-matching in the results.  Apart from the default weighting, these two options are available for reprocessing in the QA following the initial fitting.

Each photometric measurement contributes a single value to $\chi^{2}$, weighted by the fractional uncertainty of each measurement to determine the relative weighting.  The weights are initially normalized such that a photometric source with $1\%$ uncertainty has unit weight.  These are then scaled by the maximum weight over all the overlaps (i.e., the maximum number of samples in the default case) to approximately equalize the effect of the photometry and the overlaps on the solution.  The overall weighting of the photometry can also be adjusted for individual spectra as part of the QA.

For the Spitzer IRS spectra, each sample contributes separately to $\chi^{2}$, and these are included as-is, so the effective weighting is generally very high relative to the overlaps and any photometry, effectively anchoring overlapping band segments to the IRS values.

\section{SWS file header contents}\label{header}

Table \ref{tab.header} provides an example of the metadata (header) for each SWS spectral file for the version that contains both the normalized spectral segments and the original pws data, with overlaps included ({\em tdtnumber\_newpws.tbl}). The only difference to the header for the trimmed version ({\em tdtnumber\_newsws.tbl}) is that the descriptions for the pws data columns are removed.
\startlongtable
\begin{deluxetable*}{lcll}
\tablecaption{Sample Header\label{tab.header}}
\tablewidth{0pt}
\tablehead{\multicolumn{4}{l}{Untrimmed example}
}
\startdata
\textbackslash TDT      &=& 24402866            &/ Target Dedicated Time\\
\textbackslash TARGET   &=& `alpha Boo'         &/ Target name \\
\textbackslash RA       &=& 213.916542          &/ J2000 (Deg) \\
\textbackslash DEC      &=& 19.184333           &/ J2000 (Deg) \\
\textbackslash SPEED    &=& 1                   &/ SWS observing speed\\ \textbackslash OBSDATE  &=& `19960718'            &/ Observation date (YYYYMMDD)\\
\textbackslash OBSTIME  &=& `00:44:00'            &/ Observation time (UT)\\
\textbackslash PROPNAME &=& `HR5340 ALF-BOO'    &/ Name given by proposer \\
\textbackslash KSPW     &=& `1.NO'                &/ KSPW group; Kraemer+2002\\
\textbackslash PROCDATE &=& `20250117'            &/ Processing date (YYYYMMDD)\\
\textbackslash PHOTFLAG &=& T                   &/ Photometric values used in normalization\\
\textbackslash IRSFLAG  &=& T                   &/ Spitzer IRS spectrum used in normalization\\
\textbackslash \\
\multicolumn{4}{l}{\textbackslash~~  For processing details see Mizuno+2025}\\
\multicolumn{4}{l}{\textbackslash~~  Offset and scaling factors for each segment}\\
\multicolumn{4}{l}{\textbackslash~~  segment identifications: 1=1A, 2=1B, 3=1D, 4=1E, 5=2A, 6=2B,} \\
\multicolumn{4}{l}{\textbackslash~~  7=2C, 8=3A, 9=3C, 10=3D, 11=3E (not used), 12=4} \\
\textbackslash \\
\textbackslash OFFSET01 &=&  $-$426.920\\
\textbackslash OFFSET02 &=&  $-$609.510\\
\textbackslash OFFSET03 &=&   262.300\\
\textbackslash OFFSET04 &=&     0.000\\
\textbackslash OFFSET05 &=&    81.510\\
\textbackslash OFFSET06 &=&    92.410\\
\textbackslash OFFSET07 &=&    87.670\\
\textbackslash OFFSET08 &=&   $-$16.920\\
\textbackslash OFFSET09 &=&   $-$44.740\\
\textbackslash OFFSET10 &=&     0.000\\
\textbackslash OFFSET11 &=&  $-$999.000\\
\textbackslash OFFSET12 &=&     5.230\\
\textbackslash SCALE01  &=&     0.8769\\
\textbackslash SCALE02  &=&     0.8491\\
\textbackslash SCALE03  &=&     0.9462\\
\textbackslash SCALE04  &=&     1.0000\\
\textbackslash SCALE05  &=&     1.2380\\
\textbackslash SCALE06  &=&     1.3055\\
\textbackslash SCALE07  &=&     1.3104\\
\textbackslash SCALE08  &=&     1.0959\\
\textbackslash SCALE09  &=&     1.2853\\
\textbackslash SCALE10  &=&     1.0000\\
\textbackslash SCALE11  &=&  $-$999.0000\\
\textbackslash SCALE12  &=&     1.0548\\
\textbackslash\\
\multicolumn{4}{l}{\textbackslash~~   wavelength (micron)}\\
\multicolumn{4}{l}{\textbackslash~~   \underline{\hspace{0.6cm}} wavelength}\\
\multicolumn{4}{l}{\textbackslash~~   flux (Jy)}\\
\multicolumn{4}{l}{\textbackslash~~   \underline{\hspace{0.6cm}} flux density, adjusted}\\
\multicolumn{4}{l}{\textbackslash~~   error (Jy)}\\
\multicolumn{4}{l}{\textbackslash~~   \underline{\hspace{0.6cm}} uncertainty in the adjusted flux density}\\
\multicolumn{4}{l}{\textbackslash~~   segment}\\
\multicolumn{4}{l}{\textbackslash~~  \underline{\hspace{0.6cm}} segment number} \\
\multicolumn{4}{l}{\textbackslash~~  flux\_pws (Jy)}\\
\multicolumn{4}{l}{\textbackslash~~  \underline{\hspace{0.6cm}} flux density, original pws data; Sloan+2003}\\
\multicolumn{4}{l}{\textbackslash~~  error\_pws (Jy)}\\
\multicolumn{4}{l}{\textbackslash~~  \underline{\hspace{0.6cm}} uncertainty in the original flux density}\\
\multicolumn{4}{l}{\textbackslash~~  overlap\_flag}\\
\multicolumn{4}{l}{\textbackslash~~  \underline{\hspace{0.6cm}} $-$1 = no, overlapping segment number used in normalization if yes}\\
\enddata    
\end{deluxetable*}

\end{document}